\documentclass[fleqn,twoside]{article}
\usepackage{espcrc2}

\usepackage{graphicx}
\usepackage[figuresright]{rotating}

\newcommand{\AmS}{{\protect\the\textfont2
  A\kern-.1667em\lower.5ex\hbox{M}\kern-.125emS}}

\title{Gauge Theories from Dp-branes}

\author{Paolo Di Vecchia\address[MCSD]{NORDITA \\
     Blegdamsvej 17, 2100 Copenhagen \O , Denmark}}
       
\begin{document}

\begin{abstract}
In this talk we discuss the need  to introduce  a
string theory in order to obtain a consistent quantum theory of
gravity unified with gauge interactions. We then discuss some basic 
properties of string theory and the origin and the properties of the 
D(irichlet)-branes. Finally we use them for discussing the Maldacena
conjecture and its extension to non-conformal and less supersymmetric 
theories. 
\vspace{1pc}
\end{abstract}

\maketitle

\section{BEYOND THE STANDARD MODEL}
Strong, weak and electromagnetic interactions are described by the
standard model that is a gauge field theory based on the group
$SU(3)_c \otimes SU(2)_L \otimes U(1)_Y$. It has three gauge
coupling constants $g_1, g_2 $ and $g_3$ and the gauge fields are the
$8$ gluons, $W^{\pm}$, $Z^0$ and the photon. It contains two scales:
the QCD scale $\Lambda_{QCD} \sim 250\, MeV$ corresponding to the
dimension of a proton that is  about $10^{-13} cm = 1 \, Fermi$ and the
Fermi scale $\sim 250 \,GeV$ corresponding to the scale at the which the
gauge group $ SU(2)_L \otimes U(1)_Y$ is broken into $U(1)_{em}$. All
present experimental data fully agree with high precision with the
predictions of the standard model. As a consequence we can at the
moment only speculate on what will happen at higher energy and on which
additional scales we could expect. Since the running of the coupling
constants $g_i$ is entirely predictable from the low-energy particle
spectrum and quantum numbers with respect to the gauge groups of the
standard model, one can ask oneself if they have the tendency to get 
together at some higher energy. It turns out that indeed they do  at an
energy of the order of $M_{GUT} = 10^{16} \, GeV$ and this suggests that 
the three groups of the standard models may get unified at such a 
high energy~\cite{AMALDI}. In addition we know that
gravity becomes strong at the Planck mass given by:
\begin{equation}
M_{P\ell} \equiv \sqrt{\frac{\hbar c}{G_N}} = 1.2 \cdot 10^{19} GeV 
\label{planck87}
\end{equation}
This means that the standard model, although renormalizable, cannot be
a fundamental theory valid at all energies. It is only an effective
theory valid at scales $<< M_{GUT}, M_{P\ell}$. But, if this is the
case, then we
get the hierarchy problem because  we expect a Higgs
particle with a mass of the order of the cut-off corresponding in our
case to $M_{GUT}$ or $M_{P\ell}$, while we need a Higgs particle with
a mass of the order of the Fermi scale $<< M_{GUT}$ in order to break 
 $ SU(2)_L \otimes U(1)_Y$ of the standard model in  $U(1)_{em}$.
The most popular way out of this problem is to extend the standard
model to the minimal supersymmetric standard model where for each
particle of the standard model we include also its supersymmetric
partners that are required to have a mass of the order of the Fermi scale.
Actually it turns out  that in this case the three running coupling
constants meet  all at the same point corresponding to  an energy
equal to $2 \cdot 10^{16}\, GeV$~\cite{AMALDI}. 
But also the supersymmetric standard model cannot be a
fundamental theory because it does not incorporate quantum gravity and
we know that when we reach the Planck mass a classical description of
gravity is not anymore consistent. This follows from the fact
that a quantum field theory of gravity is not renormalizable. In fact 
a theory based on pointlike constituents as the case of a field theory
has already at the
classical level problems due to the short-distance or ultraviolet
divergences. These divergences are in fact already present at the
classical level in  electrodynamics~\cite{LANDAU}. That is why one
introduces the classical electron radius that is just an ultraviolet
cut-off given by:
\begin{equation}
\frac{e^2}{r_0} = mc^2 \rightarrow r_0 = \alpha \frac{\hbar}{mc} =
 \frac{1}{137}\frac{\hbar}{mc}
\label{claelera}
\end{equation}
where $\alpha$ is the fine structure constant of electromagnetism. 
In the quantum theory some of those divergences in general survive and
we must renormalize the theory in order to compare with the
experiments. But this requires  the quantum theory to be
renormalizable. Since this is not the case for quantum gravity then it
is natural, in order to construct a quantum theory of gravity, to go away
from the pointlike structure. The simplest case is that of a
one-dimensional string. It turns out that a string theory is able to
provide us a consistent quantum theory of gravity unified with gauge theories. 
Some basic elements of string theory will be discussed in the next section.

\section{STRING THEORY}
The action of the bosonic string can be constructed in analogy with that
of a spinless particle. The motion of a  spinless particle is
described by its coordinate $x^{\mu} (\tau)$ in flat Minkowski space
as a function of an arbitrary parameter $\tau$. Analogously the motion
of the bosonic string is described by its coordinate  $x^{\mu} (\tau, \sigma)$ 
as a function of two arbitrary parameters $\sigma$ and $\tau$. As the
action for a free spinless particle is proportional to the lenght of
its world-line, so the action of the bosonic string will be proportional to
the area of its world-sheet:
\begin{equation}
- mc \int \sqrt{- dx_{\mu} dx^{\mu}} \Longrightarrow -Tc \int \sqrt{-
  d\sigma_{\mu \nu} d \sigma^{\mu \nu}} 
\label{act49}
\end{equation}
where $m$ is the particle mass and $T$ is the string tension having
dimension of an energy per unit lenght.
The fact that the actions in eq.(\ref{act49}) are proportional to
geometrical objects
implies that the variables $\tau$ and $\sigma$ can be reparametrized
at will without changing the physics. This is reflected in the fact
that both actions in eq.(\ref{act49}) are invariant under
reparametrizations of the world-sheet(line) variables. In order to
quantize the system is, however, convenient to choose a covariant
gauge; the proper-time gauge in the case of the point-particle where
$\tau$ is identified with the proper-time of the particle and the
orthonormal gauge characterized by the conditions 
\begin{equation}
{\dot{x}}^2 + (x')^2 = {\dot{x}} \cdot x' =0 
\label{congau49}
\end{equation}
in the case of the bosonic string. 
In these gauges the previous actions become quadratic in the
coordinate $x^\mu$:
\begin{equation}
 - \frac{(mc)^2}{2}  \int d\tau {\dot{x}}^2 \Longrightarrow \frac{Tc}{2} \int
d\sigma d\tau  \left[ (x')^2 - {\dot{x}}^2 \right] 
\label{progau6}
\end{equation}
where ${\dot{x}}$ is the derivative with respect to $\tau$ and $x'$
that with respect to $\sigma$.

In the orthonormal gauge the string eq. of motion  becomes
linear and one gets:
\begin{equation}
\left( \frac{\partial^2}{\partial \sigma^2} - 
\frac{\partial^2}{\partial \tau^2} \right) x^{\mu} (\tau, \sigma) =0 
\label{eqmot49}
\end{equation}
that must be implemented together with eq.s (\ref{congau49}).
In addition one gets also the following boundary condition:
\begin{equation}
\frac{\partial x^{\mu}}{\partial \sigma} \cdot \delta x
|_{\sigma=0}^{\sigma -\pi} =0
\label{bounco64}
\end{equation}
where the variable $\sigma$ has been taken to vary in the interval
$(0, \pi)$. In the case of a closed string satisfying the periodicity 
condition
$x^{\mu} (\tau, \sigma) = x^{\mu} ( \tau, \sigma + \pi)$ one gets the
  following most general solution of the eq, of motion:
\[ 
x^{\mu} (\tau , \sigma ) = q^{\mu} + 2 \alpha' p^{\mu} \tau +
\]
\[
+i \frac{\sqrt{2 \alpha'}}{2}\sum_{n=1}^{\infty}\frac{1}{\sqrt{n}} 
\left[a_{n}^{\mu} {\rm e}^{-2in (\tau - \sigma)} - {a_{n}^{\dagger}}^{\mu}
 {\rm e}^{2in (\tau - \sigma)}   \right] +
\]
\begin{equation}
+i \frac{\sqrt{2 \alpha'}}{2}\sum_{n=1}^{\infty}\frac{1}{\sqrt{n}} 
\left[{\tilde{a}}_{n}^{\mu} {\rm e}^{-2in (\tau + \sigma)} -
 ({{\tilde{a}}_{n}^{\dagger}})^{\mu}
 {\rm e}^{2in (\tau - \sigma)}   \right] 
\label{sol98}
\end{equation}
where $a_n$ and ${\tilde{a}}_n$ are arbitrary parameters. In the case
of an open string the boundary condition in eq.(\ref{bounco64}) can
be satisfied by imposing at each string end-point and for each
direction one of the two conditions:
\begin{equation}
 \frac{\partial x^{\mu}}{\partial \sigma} =0~~~or~~~\delta x^{\mu} =0  
\label{ND65}
\end{equation}
The first one  corresponds to a Neumann boundary condition and
preserves translational invariance, while the second one corresponds to
a Dirichlet boundary condition and violates translational invariance.  
Imposing a Dirichlet boundary condition corresponds to require that
the string end-point is stuck on a $p$-dimensional hyperplane called 
$p$-brane, while in
the case of a Neumann boundary condition the string end-point is free 
to move and according to eq.(\ref{congau49}) it moves with the speed
of light. For
this reason a brane on which open strings can have their end-points
fixed is called a Dp-brane. We will discuss them later on. Here we
give the general solution for an open string only in the case of all
Neumann boundary conditions. It is given by:
\[
 x^{\mu} (\tau , \sigma ) = q^{\mu} + 2 \alpha' p^{\mu} \tau + 
\]
\begin{equation}
+ i \sqrt{2 \alpha'} \sum_{n=1}^{\infty}\frac{1}{\sqrt{n}} 
\left[a_{n}^{\mu} {\rm e}^{-in \tau } {a_{n}^{\dagger}}^{\mu}
 {\rm e}^{in \tau }   \right] \cos n \sigma
\label{nbc65}
\end{equation}
The quantum theory is obtained by imposing the following canonical
commutation relations:
\begin{equation}
[ a_{n}^{\mu} , a^{\dagger \nu}_{m}] = 
[{\tilde{a}}_{n}^{\mu} , {\tilde{a}}^{\dagger \nu}_{m}] = \delta_{nm}
\eta^{\mu \nu}~,~[ q^{\mu} , p^{\nu} ] = i \eta^{\mu \nu}
\label{canoco87}
\end{equation}
The spectrum of states of an open string is given by the following
expression:
\begin{equation}
\alpha' M^2 = \sum_{n=1}^{\infty} n a^{\dagger}_{n} \cdot a_{n} -1
\label{spe76}
\end{equation}
The lowest string excitation is given by the vacuum state $|0,k>$ 
with momentum $k$ corresponding to a tachyon with mass $\alpha'
M^2 =-1$, while the next one is a massless abelian vector state $
a_{1}^{\dagger\,\mu}  |0.k>$. The gauge vector field can be made
non-abelian by introducing Chan-Paton factors at the end-points of the
string. The spectrum of the closed bosonic string is given by the
following eq.:
\begin{equation}
\frac{\alpha '}{2} M^2 = \sum_{n=1}^{\infty}\left[  n a^{\dagger}_{n} \cdot
  a_{n} +\sum_{n=1}^{\infty} n {\tilde{a}}^{\dagger}_{n} \cdot {\tilde{a}}_{n}
\right] -2
\label{spe28}
\end{equation}
together with the level matching condition
\begin{equation}
\sum_{n=1}^{\infty}  n a^{\dagger}_{n} \cdot
  a_{n} = \sum_{n=1}^{\infty} n {\tilde{a}}^{\dagger}_{n}\cdot  {\tilde{a}}_{n}
\label{levma}
\end{equation}
The lowest state $|0, k>$ is again a tachyon with $\alpha' M^2 =- 4$,
while the massless states are described by the state $
a_{1}^{\dagger\, \mu} {\tilde{a}}_{1}^{\dagger \nu} |0, k>$. Its
symmetric and traceless part corresponds to the graviton, while its
trace and its antisymmetric part correspond respectively to a dilaton
and an antisymmetric  tensor (two-form potential). 

The bosonic string contains not only the particles (gauge bosons and
gravitons) that appear in the Standard model and in the Einstein's
theory of general relativity, but also their interactions. In fact it
can be shown that the  low-energy string effective action contains the
two terms:
\begin{equation}
S = \int d^D x \sqrt{-G} 
\left[ - \frac{1}{4 g^{2}_{YM}} F_{\mu \nu}^{a} F^{a\,    \mu \nu} +  
\frac{1}{2 \kappa^2} R \right]
\label{relgau67}
\end{equation}
where the Yang-Mills coupling constant and the Newton constant are
given in terms of $\alpha'$ and of the string coupling constant $g_s$ by
\begin{equation}
g^{2}_{YM} \sim g_s (\alpha')^{(D-4)/2}
\label{gym87}
\end{equation}
\begin{equation}
 2 \kappa^2 \equiv 16 \pi
G_N  \sim g_{s}^{2} (\alpha')^{(D-2)/2}
\label{form85}
\end{equation}
The quantum theory of the bosonic string is consistent with Lorentz
invariance only if the space-time dimension $D=26$. But what it makes
the bosonic string inconsistent is the presence of both the open and
closed string tachyons. In order to get rid of them one must go from
the bosonic string to the superstring. In this case one introduces
additional world-sheet degrees of freedom represented by the
world-sheet Dirac field $\psi^{\mu} (\tau, \sigma) $ corresponding to
have spin degrees of freedom along the string. There exist five
perturbatively ($g_s \rightarrow 0$) inequivalent superstring theories
that are consistent if the
space-time dimension $D=10$. They do not have any tachyon in the
spectrum if one performs the GSO projection, are space-time
supersymmetric and consistently unify gravity with gauge theories. 
They admit, however, non-perturbative solutions corresponding to
$p$-dimensional extended objects, called p-branes. If we take them into
account one can see that the five inequivalent theories are all
related to each other and the underlying unifying theory, called
M-theory, is an 11-dimensional theory that at low-energy reduces to
11-dimensional supergravity. In conclusion the long-time puzzle of the
existence of five perturbatively inequivalent string theories in $10$
dimensions has now disappeared because we have understood that at the
non-perturbative level they are all related to each other and part
of a unique $11$-dimensional M-theory.  
 
We conclude this section by observing that string theory is not in
contradiction with field theory but it is an extension of field theory as 
quantum mechanics and the theory of special relativity are an
extension of respectively classical mechanics and the galilean
mechanics in the sense that there is a limit corresponding
to sending the Planck constant $\hbar$ to zero and the speed of light $c$ to
$\infty$ where we recover respectively classical mechanics and galilean
mechanics. Analogously in the case of string theory
one can recover field theory if we send the string tension to $\infty$
or equivalently the Regge slope to zero ($\alpha' \rightarrow 0)$. In
this way one goes from string theory containing quantum gravity to
field theory containing only classical gravity.

\section{Dp-BRANES}

The massless bosonic spectrum of type II theories consists of a NS-NS
sector with a graviton $G_{\mu \nu}$, a dilaton $\phi$ and a two-form potential
$B_{\mu \nu}$ and of a R-R sector with a one-form $C_1$ and a three-form
potential $C_3$ in the case of type IIA and with a scalar
$C_0$, a two-form potential $C_2$ and a four-form potential $C_4$ in
the case of type IIB. They contain also a massless fermionic sector
consisting of two gravitinos and two dilatinos. A  one-form
potential as the electromagnetic field is coupled naturally with a
point-particle through  the well known coupling term $\int C_{\mu} d
x^{\mu} = \int C_1$. So it is natural to  think that a two-form
potential is coupled  to a
string through the coupling $\int \frac{1}{2} C_{\mu \nu} d x^{\mu}
\wedge d x^{\nu} = \int C_2$, and more generally a $(p+1)$-form
potential is coupled to a
$p$-dimensional object (called p-brane) through the interaction term
$\int C_{p+1}$. In the previous formulas we have used for convenience
the formalism of the forms where the form field is given by 
\begin{equation}
 C_{p+1} = \frac{1}{(p+1)!} C_{\mu_1 \dots \mu_{p+1}} d x^{\mu_1}
 \wedge \dots d x^{\mu_{p+1}}
\label{formfor}
\end{equation} 
The low-energy string effective action for type II theories is given
by
\[
S = \frac{1}{2 \kappa^2} \int d^{10} x \sqrt{- G} \left\{R -
 \frac{1}{2} G^{\mu \nu} \partial_{\mu} \phi \partial_{\nu} \phi \,+
\right.
\]
\begin{equation}
\left.
- \frac{1}{2 \cdot 3!} {\rm e}^{- \phi} H_{3}^2 
  -  \sum_p \frac{1}{2 \cdot (p+2)!} F_{p+2}^2 
{\rm e}^{(3-p)\phi/2}  \right\}
\label{lowene3}
\end{equation}
where the ten-dimensional Newton constant is given by $2 \kappa^2 =
 (2\pi)^7 (\alpha')^4 g_s^2$, $H_3$ is the field strenght
 corresponding to the NS-NS two-form potential and $F_{p+2}$ is the
 field  strenght
 corresponding to the potential $C_{p+1}$. The action in
 eq.(\ref{lowene3}) contains
additional terms that are not important for our considerations.

Dp-branes are non-perturbative solutions of the classical equations of
motion obtained from the previous low-energy string effective action
in which at least one of the R-R fields is turned on. One starts from
the following ansatz for a Dp-brane solution:
\[
(ds)^2 \equiv G_{\mu \nu} dx^{\mu} dx^{\nu} 
= [ H(r) ]^{(p-7)/8}
\eta_{\alpha \beta} dx^{\alpha} dx^{\beta}+
\]
\begin{equation}
 + [H (r)]^{(p+1)/8}
\delta_{ij} dx^{i} dx^{j}
\label{metric2}
\end{equation}
for the metric, where we have called the coordinates $\mu, \nu$ with
the indices $(\alpha, \beta)$ for the directions along the world-volume of
the brane and with the indices $(i,j)$ for those transverse to the
brane. In particular $\alpha, \beta = 0, \dots p$ and $i,j = (p+1),
\dots 9$. In addition to the metric we have also the dilaton and a R-R
field turned on and they are given by:
\begin{equation}
{\rm e}^{-(\phi - \phi_0)} = [H(r)]^{(p-3)/4}~~,~~C_{01 \dots p} =
\frac{1}{H} -1
\label{dilrr}
\end{equation}
The harmonic function $H$ is given by:
\begin{equation}
H(r) = 1 + \frac{K_p N}{r^{7-p}}~~~~,~~~r^2 = \sum_i x_i^2
\label{harmonic}
\end{equation}
where
\begin{equation}
K_p = \frac{2 \kappa^2 \tau_p}{(7-p) \Omega_{8-p}}~~~,~~~\Omega_q =
  \frac{2 \pi^{(q+1)/2}}{\Gamma ( \frac{q+1}{2} )}
\label{kp}
\end{equation}
and $\Omega_q$ is the volume of the $q$-dimensional sphere. The
tension of the brane, that is equal to its mass per unit of p-volume,
is given by
\begin{equation}
\frac{M}{p-volume} = \int d^{9-p} x \, \, \theta_{00} = \tau_p N
\label{tens}
\end{equation}
while its charge is given by
\[
\frac{1}{\sqrt{2} \kappa} \int_{S^{8-p}} {\rm e}^{- ((p-3) \phi/2}
{}^*F_{8-p} = 
\]
\begin{equation}
 = \sqrt{2} \kappa \tau_p N \equiv \mu_p N
\label{charge}
\end{equation}
In terms of the string parameters $g_s$ and $\alpha'$ they are given
by
\begin{equation}
\tau_p = \frac{ (2 \pi \sqrt{\alpha'})^{-p}}{ g_s \sqrt{\alpha '}}~~,~~
\mu_p = \sqrt{2 \pi} \, (2 \pi \sqrt{\alpha '})^{3-p}
\label{taumu}
\end{equation}

In string theory the Dp-branes are characterized by the fact that
open strings have their end-points attached to their world-volume. The
spectrum of open strings having  their end-points attached to the
world-volume of a Dp-brane (i.e. satisfying Neumann boundary
conditions on the directions along the world-volume of the brane and
Dirichlet boundary conditions along the directions transverse to the
brane) is given by the following formula:
\begin{equation}
\alpha' k_{||}^2 + \sum_{n=1}^{\infty} n a^{\dagger}_{n} \cdot a_n +
\sum_t t \psi^{\dagger}_{t} \cdot \psi_t -a =0
\label{speope}
\end{equation}
where $a = \frac{1}{2} [0]$ in the NS [R] sector and $k_{||}$ is the
momentum of the string parallel to the brane. In particular the
massless states in the NS sector are given by $ ( \psi_{1/2}^{\alpha}, 
\psi_{1/2}^{i}) | 0, k >$ corresponding to a gauge boson $A_\alpha$  and
to $(9-p)$  Higgs scalars $\Phi^i$ related to the translational modes
of the brane along the directions transverse to its
world-volume. These gauge and scalar fields living on the world-volume
of a Dp-brane become non-abelian transforming all of them according to
the adjoint representation of the gauge group if instead of a single
Dp-brane we have a bunch of $N$ coincident Dp-branes. In this case in
fact we get  $N^2$ massless states
corresponding to the fact that the open strings can have their
end-points on each of the $N$ branes. In conclusion, while a single
Dp-brane will have an abelian  gauge field and $(9-p)$ Higgs fields
living on its world-volume, a bunch of $N$ coincident Dp-branes will
have a non-abelian gauge field and $(9-p)$ Higgs fields all
transforming according to the adjoint representation of $U(N)$.     

The low-energy dynamics of a Dp-brane is described by the Born-Infeld
action that is given by:
\[
S_{BI} = - \tau_p \int d^{p+1} x \,\,{\rm e}^{-(3-p)\phi /4} \times
\]
\[
\sqrt{- \det \left[G_{\alpha \beta} + {\rm e}^{-\phi/2}
    \left(B_{\alpha \beta} + 2 \pi \alpha' F_{\alpha \beta} \right)
    \right]} \,+
\]
\begin{equation}
+ \tau_p \int_{V_{p+1}} \sum_n C_{n} {\rm e}^{(2\pi \alpha' F +B)}
\label{BI12}
\end{equation}
It contains the coordinates of the brane represented by the gauge
field $A_{\alpha}$ living on the brane with field strenght 
$F_{\alpha \beta}$ and by the transverse brane coordinates related to
the Higgs fields by the relation $ x^i \equiv 2 \pi \alpha'
\Phi^i$. They correspond to the massless open string excitations  of the
NS sector. This means that the dynamics of a brane is determined by
the open strings having their end-points on the brane. The Born-Infeld
action in eq.(\ref{BI12})
contains also the bulk fields (i.e. fields living in the entire
ten-dimensional space and not just on the brane) corresponding to the massless 
closed string excitations of the NS-NS and R-R sectors. What appears
in eq.(\ref{BI12}) is actually their pullback on the brane defined by
\begin{equation}
G_{\alpha \beta} = G_{\mu \nu} \partial_{\alpha} x^{\mu}
\partial_{\beta} 
x^{\nu}~~,~~ B_{\alpha \beta} = B_{\mu \nu} \partial_{\alpha} x^{\mu}
\partial_{\beta} 
x^{\nu} 
\label{pullba}
\end{equation}
with a similar expression for the R-R fields.
In the case of a system of $N$ coincident branes the Born-Infeld
action gets modified by the fact that the coordinates of the branes
become non-abelian fields and the brane tension $\tau_p$ gets
multiplied with a factor $N$. The complete expression of the
non-abelian Born-Infeld action is not yet known. But for our purpose
it is sufficient to consider the non-abelian extension given in
Ref.~\cite{TSE} where the symmetrized trace is introduced.

On the one hand a system of $N$ Dp-branes is a classical solution of
the low-energy string effective action whose low-energy dynamics is
described by the Born-Infeld action. In particular it can be shown that
a system of $N$ coincident Dp-branes is a BPS state preserving $1/2$
supersymmetry (corresponding to $16$ preserved supersymmetries) and as 
a consequence they are not interacting. This can be easily seen by 
plugging  the classical solution given in eq.s (\ref{metric2}) and 
(\ref{dilrr}) in the Born-Infeld action in eq.(\ref{BI12}). In fact if
we do that neglecting the coordinates of the brane we get
\[
\tau_p \int d^{p+1} x \left\{ - H^{ [(p- 7)(p+1) - (p-3)^2]/16} + \right.
\]
\begin{equation}
+ \left. \frac{1}{H} -1 \right\}  = -\tau_{p} \int d^{p+1} x
\label{noforce18}
\end{equation}
that is independent on the distance $r$ between the brane probe
described by the Born-Infeld action and the system of $N$ coincident
branes described by the classical solution. 

On the other hand a system of $N$ Dp-branes has a $U(N)$ gauge theory living
on its world-volume with $16$ supersymmetries corresponding, in the
case of $p=3$, to ${\cal{N}}=4$ super Yang-Mills in four dimensions
that is a conformal invariant theory with vanishing
$\beta$-function. Its Lagrangian can be obtained by expanding the
first term of the Born-Infeld action up to the quadratic order in the
gauge fields living on the brane. Neglecting the term independent from
the gauge fields that we have already computed in eq.(\ref{noforce18})
we get the following Lagrangian:
\begin{equation}
L = \frac{1}{g_{YM}^{2}} \left[- \frac{1}{4} F_{\alpha \beta} F^{\alpha
    \beta}  + \frac{1}{2} \partial_\alpha \Phi^i \partial^{\alpha}
    \Phi^i  \right] + \dots
\label{YMax45}
\end{equation}
where the gauge coupling constant is a constant given by:
\begin{equation}
g_{YM}^{2} = \frac{2}{\tau_p (2 \pi \alpha ')^2} = \frac{2 g_s
  \sqrt{\alpha'} ( 2 \pi \sqrt{\alpha'})^p}{(2 \pi \alpha')^2}
\label{couco45}
\end{equation}
In particular for $p=3$ we get $ g_{YM}^{2} = 4 \pi g_s$. The action in 
eq.(\ref{YMax45}) corresponds to the dimensional reduction of
the ${\cal{N}}=1$ super Yang-Mills in ten dimensions to $(p+1)$
dimensions.

The previous considerations imply that the low-energy dynamics of
branes can be used to determine the properties of gauge theories and 
viceversa. 
 
Let us analize now what is the range of validity of the classical
solution given in  eq.s (\ref{metric2}) and (\ref{dilrr}) restricting
to the most interesting case of $p=3$. In order to
be able to use classical gravity we need, on the one hand, to neglect
closed string loops and, on the other hand, to restrict the curvature
of the solution to be small. The first condition implies that $g_s <<1$.
But, if we kook at eq.(\ref{harmonic}) and we keep $N$ small, the
condition $g_s <<1$ is equivalent to large values of $r$ where the
metric is almost Minkowski. Therefore in this regime the curvature is
also small. This means that the classical solution provides a
consistent description of the brane for large values of $r$. On the
other hand, if we take the number of branes to be large in such a way
that $N g_s$ is not necessarily small, then we can go to the
near-horizon ($r \rightarrow 0$) limit of the classical solution and
we can ask ourselves: what is the value of $N g_s$ corresponding to a
small curvature? It turns out that this happens for strong values of
the 't Hooft coupling $N g_s >>1$. This can formally be seen by taking
the low-energy ($\alpha' \rightarrow 0$) and the near-horizon ($r
\rightarrow 0$) limit of the D3-brane classical solution keeping the
quantity $U \equiv \frac{r}{\alpha'}$ fixed and corresponds to neglect
the term $1$ in eq.(\ref{harmonic}) as follows from 
\[
H = 1+ \frac{4 \pi g_s (\alpha')^2 N}{r^4} = 
\]
\begin{equation}
 = 1 + \frac{4 \pi g_s
  N}{(\alpha')^2 U^4} \rightarrow  \frac{4 \pi g_s
  N}{(\alpha')^2 U^4}
\label{nearho45}
\end{equation}
In this limit we get the metric of $AdS_5 \times S^5$:
\begin{equation}
\frac{ds^{2}}{\alpha'} \rightarrow \frac{U^2}{b^{2}/\alpha'}dx_{3+1}^2 +
\frac{b^2/\alpha'}{U^2}dU^2 + \frac{b^2}{\alpha'} d \Omega_{5}^{2} 
\label{nearho28}
\end{equation}
where the radii of $AdS_5$ and of $S^5$ are equal and given by
\begin{equation}
R^{2}_{AdS_5} = R_{S^5} = b^2 = \sqrt{4 \pi g_s N} \alpha'
\label{rad}
\end{equation}
These formulas imply that the condition of small curvature requires
that $g_s N >>1$. In conclusion the classical solution provides a good
description of the brane at  large distance ($r \rightarrow \infty$)
if $N g_s <<1$ and in the near-horizon limit ($r \rightarrow 0$) if
$Ng_s >>1$.

In the next section starting from the previous considerations we will
formulate the Maldacena conjecture for ${\cal{N}}=4$ super Yang-Mills theory.

\section{MALDACENA CONJECTURE}

In the previous section we have seen that a D3-brane is described at
low-energy either by the Born-Infeld action that for $\alpha'
\rightarrow 0$ reduces to ${\cal{N}}=4$ super Yang-Mills or by a
classical  solution of supergravity equations. They are different but 
equivalent ways of
describing a D3-brane. In the following we will use these two
different, but equivalent descriptions of a D3-brane for arriving at
the Maldacena conjecture~\cite{MALDA} stating that ${\cal{N}}=4$ 
super Yang-Mills
is equivalent to ten-dimensional type IIB string theory compactified
on  $AdS_5 \times S^5$.  In particular, if we consider the low-energy
limit ($\alpha' \rightarrow 0$) of the Born-Infeld action, it consists
of a brane action corresponding to that  of four-dimensional
${\cal{N}}=4$ super Yang-Mills and by a term describing the
interaction of the brane with the bulk fields. However, in the limit
of $\alpha ' \rightarrow 0 $ the interaction term, being proportional
to $\kappa \sim (\alpha')^2$, is vanishing. In this limit also the
bulk fields are not interacting. Therefore from the point of view of
the Born-Infeld action in the low-energy limit we get ${\cal{N}}=4$
super Yang-Mills plus free gravitons or more in general free bulk fields.  

On the other hand, if we look at the classical solution given in 
eq.(\ref{metric2}) we see that it interpolates between flat Minkowski
space  ($r \rightarrow \infty$) and a long throat in the
near-horizon limit ($r \rightarrow 0$).
If we have sufficiently soft gravitons (i.e. gravitons with wave lenght much
bigger than the radius of the throat $b$) outside the throat they cannot 
interact with the excitations far down in the throat as it is confirmed by
the fact that their absorption cross-section  is vanishing at low 
energy~\cite{KLEBA1,KLEBA2}.  
On the other hand a string excitation far down inside 
the throat, although its proper energy (the energy measured in the reference 
frame instantaneously at rest at $r$) diverges at low energy ($\alpha ' 
\rightarrow 0)$, being proportional to $ E_p 
\sim 1/\sqrt{\alpha '}$, is not negligible because its energy measured in 
the frame of reference where the time is the one appearing in the first term
of the r.h.s. of eq.(\ref{nearho28})  is given by:
\begin{equation}
E_t \sim \frac{r}{b} E_p \sim \frac{r}{b \sqrt{\alpha'}} \sim \frac{r}{\alpha'}
= U
\label{et}
\end{equation}
that is kept fixed in the limit $\alpha ' \rightarrow 0$. Therefore from the
point of view of the classical solution we are left with free gravitons and
all the string excitations living far down inside the throat that are 
described by type IIB string theory compactified on $AdS_5 \times
S^5$. 

By comparing this
result with the one obtained from the Born-Infeld action Maldacena 
 has formulated the conjecture that ${\cal{N}} =4$ super Yang-Mills is
equivalent to type IIB string theory compactified on $AdS_5 \times S^5$.
The precise relation between the parameters of the gauge and string theories
is given in eq.(\ref{rad}), where $N$ is equal to the number of colours in the
gauge theory and to the flux of the $5$-form field strenght in the supergravity
solution. Since the classical solution in eq.(\ref{nearho28}) is a good 
approximation when the radii of $AdS_5$ and $S^5$ are very big 
\begin{equation}
\frac{b^2}{\alpha'} >> 1 \Longrightarrow Ng_{YM}^{2} \equiv \lambda >>1~,
\label{bigrad}
\end{equation}
in the strong coupling limit of the gauge theory we can restrict ourselves
to the type IIB supergravity compactified on $AdS_5 \otimes S^5$.
 
In conclusion, according to the Maldacena conjecture, classical supergravity
is a good approximation if $\lambda >>1$, while in the 't Hooft limit in which
$\lambda$ is kept fixed for $N \rightarrow \infty$ classical string theory
is a good approximation for  ${\cal{N}}=4$ super Yang-Mills. In the 't Hooft
limit in fact string loop corrections are negligible ($g_s << 1$) 
as it follows from the equation: $ \lambda = 4 \pi g_s N$ for $\lambda$ fixed
and $N \rightarrow \infty$. Finally Yang-Mills
perturbation theory is a good approximation when $\lambda <<1$.

The strongest evidence for the validity of the Maldacena conjecture comes from
the fact that both ${\cal{N}}=4$ super Yang-Mills and type IIB string 
compactified on $AdS_5 \otimes S^5$ have the same symmetries. They are, in 
fact, both invariant under $32$ supersymmetries, under the conformal group
$O(4,2)$, corresponding to the isometries of $AdS_5$, under the $R$-symmetry
group $SU(4)$, corresponding to the isometries of $S^5$ and under the 
Montonen-Olive duality~\cite{MONTOLI} based on the group $SL(2,Z)$. 

The validity of the Maldacena conjecture has by now been confirmed by
many checks and this is the first time that a 
string theory is recognized to come out from a gauge theory. In particular
it is important to stress that this does not contradict the fact  
that a string theory contains gravity while the gauge theory does not,
because in this case the two theories live in different spaces: IIB string 
theory lives on 
$AdS_5 \otimes S_5$, while ${\cal{N}} =4$ super Yang-Mills lives in our
four-dimensional Minkowski space. 
A new puzzle,
however, arises in this case because we usually connect a string theory
with a confining gauge theory, while instead ${\cal{N}}=4$ super Yang-Mills
is a conformal invariant theory and therefore is in the Coulomb and not in
the confining phase.

\section{NONCONFORMAL GAUGE THEORIES}
In the previous section we have seen that IIB string theory
compactified on $AdS_5 \times S^5$ (IIB supergravity
for large values of the 't Hooft coupling) can be used to study the
properties of ${\cal{N}}=4$ super Yang-Mills. It is desiderable to
extend the previous analysis to less supersymmetric and non conformal
gauge theories. Many different attempts have been tried to describe
non-conformal and less supersymmetric theories using supergravity
solutions. In this section we limit ourselves to discuss how to use the
fractional branes for studying the properties of ${\cal{N}}=2$ super
Yang-Mills. 

Let us consider type IIB string theory on the background $R^{1,5}
\times R^4 /Z_2 $ consisting of  $6$-dimensional Minkowski space
times the orbifold $R^4 /Z_2$ where $Z_2$ acts on the
orbifolded components by changing their sign: $ x^i \rightarrow - x^i$
where $i=6,7,8,9$. Such a background breaks $1/2$ supersymmetry with
respect to  ten-dimensional Minkowski space and on the world-volume
of a D3-brane that breaks an additional $1/2$ supersymmetry, one gets a
supersymmetric theory with $8$ instead of $16$ charges as in the case
of ${\cal{N}} =4$ super Yang-Mills. On the other
hand in the case of an orbifold one has a more general type of branes,
called fractional Dp-branes. They are characterized by the fact that they
are stuck at the orbifold fixed point ($x^6 = x^7 = x^8 = x^9 =0$),
have a charge and tension that are a fraction of those of a normal
brane (in the case of a $Z_2$ orbifold this fraction is just $1/2$)
and can be seen as D(p+2)-branes wrapped on  vanishing exceptional
cycles located at the orbifold fixed points.  

Since we want to study four-dimensional gauge theories let us consider
a fractional D3-brane of the orbifold $R^4/Z_2$, let us 
write its world-volume action and let us use it to determine
the corresponding IIB supergravity classical solution. Such a D3-brane
is coupled to the untwisted bulk fields corresponding to the metric
$G_{\mu \nu}$ and the $4$-form potential $C_4$ as a normal
D3-brane. In addition it is coupled to the two twisted fields $b,c$
obtained by wrapping the two $2$-form potentials $B_2$ and $C_2$ on
the vanishing exceptional cycle:
\begin{equation} 
B_2 = b \omega_2~~,~~C_2 = c \omega_2
\label{wra98}
\end{equation}  
A fractional D3-brane is described by the following Born-Infeld
action~\cite{BERTOLINI}:
\[
S_{BI}= \frac{\tau_3}{ 4 \pi^2 \alpha'} \left[ - \int d^4 x 
  \,\, b \,\, \times \right.
\]
\[ 
\sqrt{- \det \left( G_{\alpha \beta} + 2 \pi \alpha'  F_{\alpha \beta} 
\right)} 
\]
\begin{equation} 
+ \left. \int C_4 b + \int {\cal{A}}_4 \right] + \frac{c}{2 \pi g_s}
( \frac{1}{32 \pi^2} \int d^4 x F_{\alpha \beta} {\tilde{F}}^{\alpha \beta})
\label{BI765}
\end{equation}  
where ${\cal{A}}_4$ is the Hodge dual of $c$.

The classical supergravity solution corresponding to a system of $M$ 
fractional
D3-branes is given by the following expressions for the untwisted fields:
\begin{equation} 
ds^2 = H^{-1/2} \eta_{\alpha \beta} d x^{\alpha} d x^{\beta} + H^{1/2}
  \delta_{ij} dx^i d x^j
\label{metri74}
\end{equation}  
and
\begin{equation} 
{\tilde{F}}_5 = H^{-1} d x^0 \dots d x^3 + {}^{*}d ( H^{-1} d x^0 \dots d x^3)
\label{c498}
\end{equation}  
where the function $H$ satisfies the  equation~\cite{BERTOLINI,POLCHI}:
\[
- \partial_i \partial^i H (r, \rho ) = 2 \kappa^{2}
\tau_3 M \delta (x^4) \dots \delta ( x^9 ) +
\]
\begin{equation} 
+\frac{(4 \pi \alpha' g_s  M)^2}{\rho^2} \delta (x^6 ) \dots \delta (
x^9 ) 
\label{heq65}
\end{equation}  
with $r^2 = \sum_{i=4}^{9} x_{i}^{2}$ and $ \rho^2 = (x^4 )^2 +
(x^5)^2$. 

The twisted fields are only a function of the coordinates $x^4$ and
$x^5$ and are given by~\cite{BERTOLINI,POLCHI}
\begin{equation} 
\gamma (z) \equiv c +i b = 2 i \pi \alpha' g_s \left[\frac{\pi}{g_s} +
2 M \log \frac{z}{\epsilon} \right]
\label{gamma43}
\end{equation}  
that implies
\begin{equation} 
b = \frac{(2 \pi \sqrt{\alpha'})^2}{2} + 4 \pi \alpha' M g_s \log 
\frac{\rho}{\epsilon}
\label{b65}
\end{equation}  
and 
\begin{equation}   
c = - 4 \pi \alpha' M g_s \theta~~~,~~~z \equiv x^4 + i x^5 = \rho
{\rm e}^{i \theta}
\label{c65}
\end{equation}

It can be seen that the metric has a naked singularity of the repulson
type (the gravitational force goes to zero and becomes repulsive) at
short distance. On the other hand if we plug the classical solution
given above in the Born-Infeld action of a probe fractional D3-brane
we see that its tension becomes negative at a distance (called
enhan{\c{c}}on) bigger than the
one corresponding to the repulson singularity. This means that the
classical solution makes sense only for distances bigger than the
enhan{\c{c}}on. When we insert the classical solution in the probe action we
find that the gauge theory living on the brane has a running coupling
constant given by:
\begin{equation} 
\frac{1}{g_{YM}^{2} (\rho)} = \frac{1}{g_{YM}^{2} (\epsilon)} +
\frac{M}{4 \pi^2} \log \frac{\rho}{\epsilon}
\label{run64}
\end{equation}  
and a $\theta_{YM}$ angle given by:
\begin{equation} 
\theta_{YM} =2 M \theta~~~~,~~~~g^{2}_{YM} (\epsilon) \equiv 8 \pi g_s
\label{theta45}
\end{equation}  
where $\rho$ that originally is the distance between the brane probe
and the branes that generate the classical solution becomes the
renormalization group scale of the gauge theory living on the
brane. The previous analysis provides a geometrical interpretation of
the value of $\mu$ where the gauge coupling constant becomes infinite
corresponding in gauge theory to $\Lambda_{QCD}$ (the dimensional
constant generated by dimensional transmutation in gauge theories) and
to the enhan{\c{c}}on (scale where the brane probe becomes tensionless) in
the brane dynamics. From eq.(\ref{run64}) one can compute the 
$\beta$-function of the gauge theory that is given by~\cite{BERTOLINI}:
\begin{equation} 
\beta (g_{YM}) = - \frac{M g_{YM}^{3}}{8 \pi^2}
\label{beta65}
\end{equation}  
and that is the $\beta$-function of ${\cal{N}}=2$ super Yang-Mills.
We see that the classical solution describes the perturbative
properties of  ${\cal{N}}=2$ super Yang-Mills, but fails to reproduce
the non-perturbative instanton contribution.

\end{document}